\documentclass[aps,prl,reprint,showpacs,superscriptaddress]{revtex4-2}

\usepackage{etoolbox}
\usepackage{graphicx,amssymb,bm,amsfonts,amsmath,graphics,color,epstopdf}
\usepackage[bookmarks=false,pdfstartview=FitH,colorlinks=true,citecolor=blue,linkcolor=blue]{hyperref}
\usepackage[version=3]{mhchem}
\newcommand {\refeq}[1]{(\ref{#1})}

\begin{document}


\title{Hidden Silicon-Vacancy Centers in Diamond}
\author{Christopher L.\ Smallwood}
\affiliation{Department of Physics, University of Michigan, Ann Arbor, MI 48109, USA}
\affiliation{Department of Physics and Astronomy, San Jos\'e State University, San Jose, CA 95192, USA}
\author{Ronald Ulbricht}
\affiliation{Max Planck Institute for Polymer Research, Ackermannweg 10, 55128 Mainz, Germany}
\author{Matthew W.\ Day}
\affiliation{Department of Physics, University of Michigan, Ann Arbor, MI 48109, USA}
\author{Tim Schr\"oder}
\affiliation{Department of Electrical Engineering and Computer Science, Massachusetts Institute of Technology, Cambridge, MA 02138, USA}
\affiliation{Department of Physics, Humboldt-Universit\"at zu Berlin, Newtonstra{\ss}e 15, 12489 Berlin, Germany}
\author{Kelsey M.\ Bates}
\affiliation{Department of Physics, University of Michigan, Ann Arbor, MI 48109, USA}
\author{Travis M.\ Autry}
\altaffiliation[Present address: ]{HRL, Malibu, CA 90265, USA}
\affiliation{JILA, University of Colorado \& National Institute of Standards and Technology, Boulder, CO 80309, USA}
\affiliation{Department of Physics, University of Colorado, Boulder, CO 80309, USA}
\author{Geoffrey Diederich}
\altaffiliation[Present address: ]{Department of Physics, University of Washington, Seattle, WA 98195, USA}
\affiliation{Department of Physics and Astronomy, University of Denver, Denver, CO 80208, USA}
\author{Edward Bielejec}
\affiliation{Sandia National Laboratories, Albuquerque, NM 87185, USA}
\author{Mark E. Siemens}
\affiliation{Department of Physics and Astronomy, University of Denver, Denver, CO 80208, USA}
\author{Steven T.\ Cundiff}
\email[Email: ]{cundiff@umich.edu}
\affiliation{Department of Physics, University of Michigan, Ann Arbor, MI 48109, USA}
\date {\today}

\begin{abstract}
We characterize a high-density sample of negatively charged silicon-vacancy (SiV$^-$) centers in diamond using collinear optical multidimensional coherent spectroscopy. By comparing the results of complementary signal detection schemes, we identify a hidden population of \ce{SiV^-} centers that is not typically observed in photoluminescence, and which exhibits significant spectral inhomogeneity and extended electronic $T_2$ times. The phenomenon is likely caused by strain, indicating a potential mechanism for controlling electric coherence in color-center-based quantum devices.
\end{abstract}


\maketitle


%

Color centers in diamond are point defects within the diamond host lattice that absorb and emit visible or near-infrared light. Such defects have drawn attention recently as potential hardware elements in quantum networks and devices \cite{Awschalom2018,Bradac2019}, due in part to the protective influence of diamond's wide bandgap and weak magnetic susceptibility, and in part to the technology available for manipulating and detecting light at these photon energies. Among the most promising defects are negatively charged silicon-vacancy (\ce{SiV^-}) centers. In contrast to the more heavily studied nitrogen-vacancy (NV) centers, \ce{SiV^-} centers exhibit an inversion-symmetric $D_{3d}$ point-group structure \cite{Hepp2014,Rogers2014} that shields them against first-order Stark shifts and endows them with sharp optical absorption and emission lines. Aided by these properties, \ce{SiV^-} centers have proven conducive to both coherent control \cite{Rogers2014a,Pingault2014,Becker2016,Zhang2017} and the manipulation and generation of indistinguishable photons \cite{Sipahigil2014,Schroeder2017}.

Despite these advantages, challenges still hinder the development of \ce{SiV^-} center devices. For example, spin coherence times are in most cases too short for practical applications with the exception of measurements at sub-Kelvin temperatures \cite{Sukachev2017,Becker2018a} and of the neutral \ce{SiV}-center variant \cite{Green2017,Rose2018}. Moreover, the overall quantum yield of \ce{SiV^-} centers is low \cite{Turukhin1996,Neu2012}, suffering from non-radiative decay channels \cite{Neu2011,Gali2013,Jahnke2015} that are a potential problem for single-photon devices. Separately, there are a number of open questions and engineering opportunities related to the effects of strain on \ce{SiV^-} centers \cite{Meesala2018,Lindner2018}, and more broadly to the way that ensembles of color centers behave in close proximity \cite{Evans2018} and at large densities \cite{Bradac2017,Venkatesh2018,Kucsko2018}. A few of these studies indicate a correspondence between strain and electronic dephasing rates \cite{Lindner2018}. On all these fronts, the exceptional sensitivity of nonlinear optical techniques \cite{Becker2016,Ulbricht2016,Arend2016,Zhang2017,Ulbricht2018,Weinzetl2019} holds promise for elucidating \ce{SiV^-} center properties.

\begin{figure*}[tb]\centering\includegraphics[width=7in]{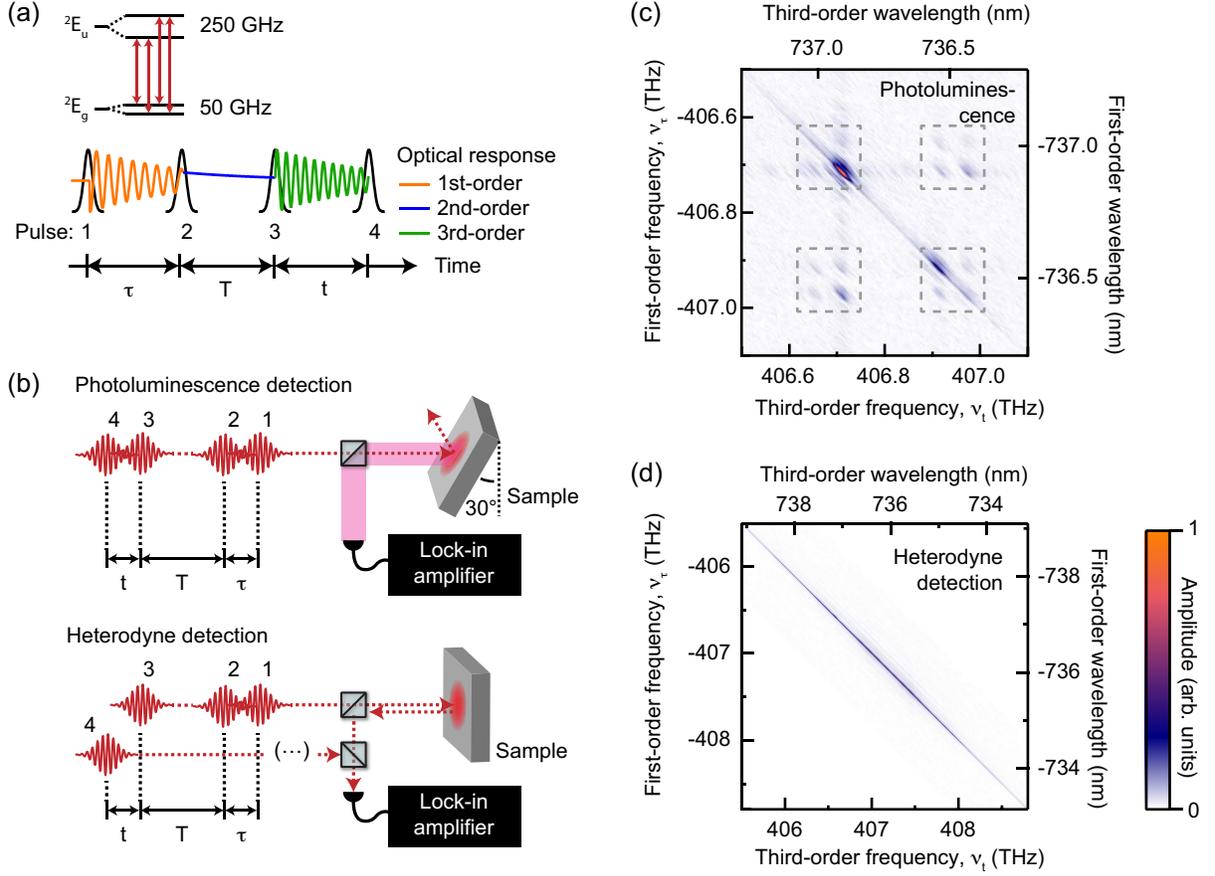}
\caption{\label{fig1}Comparison between PL-detected and heterodyne-detected multidimensional coherent spectroscopy (MDCS) measurements of an ensemble of \ce{SiV^-} centers in diamond.
{\bf(a)} Temporal illustration of relevant interaction pathways and probed electronic coherences.
{\bf(b)} Schematic illustrations of signal collection techniques.
{\bf(c)} PL-detected rephasing spectrum.
{\bf(d)} Heterodyne-detected rephasing spectrum.
}
\end{figure*}

In this Letter, we report measurements using collinear multidimensional coherent spectroscopy (MDCS) of the optical transitions in a high-density ensemble of \ce{SiV^-} centers in bulk single-crystal diamond. By comparing photoluminescence (PL)-based and heterodyne-detection-based MDCS signal collection schemes, we selectively distinguish between luminescing and nonluminescing color centers, observing a large population of \ce{SiV^-} transitions that are typically hidden (i.e., not observed) under PL detection, and which have more than 60 times as much inhomogeneous spectral broadening as the population of PL-emitting ``bright" states. A detailed comparison of homogeneous dephasing rates reveals longer $T_{2}$ optical coherence times for the hidden population as compared to the bright population, indicating that electronic dephasing interactions in the hidden population are diminished. Finally, by characterizing the amount of inhomogeneous broadening, we argue that the most likely source of the inhomogeneity and extended $T_{2}$ times of the hidden population is inhomogeneous strain, and discuss mechanisms by which strain and electronic dephasing might be linked. The results exemplify the power of MDCS as a tool for characterizing color-center materials, and they inform the development of strain-utilizing color-center devices and applications.

The experimental layout is illustrated in Fig.\ \ref{fig1}. We exposed the sample to a series of laser pulses (bandwidth $\approx$ 7.5 nm, $\lambda_{\textrm{peak}}\approx 737$ nm, repetition rate = 76 MHz) that combined to generate nonlinear polarization and excited-state-electron population responses \cite{sm}. We collected rephasing spectra \cite{HammZanni,Smallwood2018} in which a first-order excitation interaction was correlated with a third-order evolution or emission interaction [Fig.~\ref{fig1}(a)] to produce photon echoes as a function of inter-pulse time delays $\tau$ and $t$. Fourier-transforming the result generates spectra consisting of two-dimensional resonance peaks plotted against interaction frequencies $\nu_\tau$ and $\nu_t$. Measurements were performed at 10 K and a waiting time $T \approx 0.5$ ps.

An advantage of this setup over other types of MDCS (including a previous study on NV centers in diamond \cite{Huxter2013}) is that the excitation beams are collinear \cite{Tian2003,Langbein2006,Tekavec2007,Nardin2013,Martin2018}. The arrangement facilitates comparisons between complementary signal detection schemes, as shown in Fig.~\ref{fig1}(b). We isolated the nonlinear response by tagging each of the four laser pulses with frequency offsets $\nu_1$, $\nu_2$, $\nu_3$, and $\nu_4$, which led to a radio-frequency beatnote at $\nu_{sig} = -\nu_1 + \nu_2 + \nu_3 - \nu_4$ that we selected with a lock-in amplifier. In one scheme, we extracted the signal by directing all four pulses onto the sample and measuring the resulting modulation in PL intensity \cite{Tian2003,Tekavec2007}. In another scheme, we used the fourth pulse as a local oscillator to heterodyne-detect a coherent four-wave-mixing signal emitted in reflection by the interaction of the sample with the first three pulses \cite{Langbein2006,Martin2018}. The first scheme selects out bright color centers with a propensity for radiative emission while the second scheme is more sensitive to the ensemble as a whole. An analogous comparison in the linear regime would be one between photoluminescence excitation spectroscopy and absorption.

Figure \ref{fig1}(c) shows an MDCS plot for the PL detection scheme. The energy-level structure of an \ce{SiV^-} color center contains spin-orbit-split $^2E_g$ ground and $^2E_u$ excited states [see Fig.\ \ref{fig1}(a)], which give rise in Fig.~\ref{fig1}(c) to optical transitions at 406.654 THz, 406.713 THz, 406.915 THz, and 406.974 THz. (The $y$-axis values are negative because the first-order interaction is conjugate relative to the third-order interaction in a rephasing pulse sequence.) Such transitions can be observed using traditional PL \cite{Clark1995,Hepp2014,Rogers2014}, but the MDCS measurement is richer. Within the figure's dashed gray boxes, for example, there are direct peaks visible at $\nu_\tau = -\nu_t$ indicating ground-state bleaching and stimulated emission effects, and there are cross peaks at $\nu_\tau \neq -\nu_t$ indicating the presence of coherent coupling. Beyond this, the elongated resonance features offer a window into the intricacies of dephasing. The features' ``inhomogeneous" linewidths (parallel to the $\nu_\tau = -\nu_t$ line) give a measurement, in the language of magnetic resonance \cite{ErnstT2}, of the $T_2^*$ dephasing times associated with the overall ensemble. Their ``homogeneous" linewidths (perpendicular to the $\nu_\tau = -\nu_t$ line) give a frequency-group-specific measurement connected to single-particle dephasing times $T_2$.

\begin{figure}[tb]\centering\includegraphics[width=3.4in]{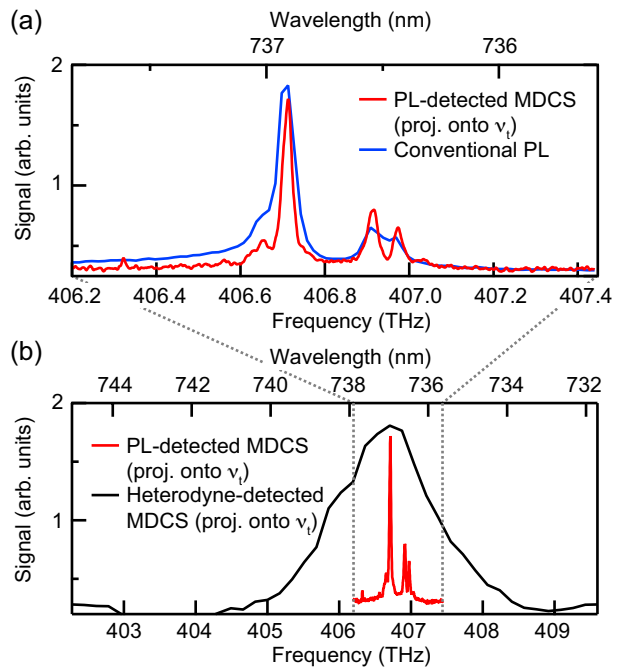}
\caption{\label{fig2}
Inhomogeneous distributions of \ce{SiV^-} centers in PL-detected and heterodyne-detected MDCS spectra, where traces have been extracted by taking the projection of the amplitude of rephasing spectra onto the $\nu_t$ axis.
{\bf(a)} PL-detected MDCS and traditional PL. The data have been scaled and vertically offset to facilitate lineshape comparisons.
{\bf(b)} PL-detected MDCS and heterodyne-detected MDCS.
}
\end{figure}

More striking than the characteristics of the PL-detected plot in isolation, however, are the dramatic differences between the PL-detected spectrum of Fig.\ \ref{fig1}(c) and the heterodyne-detected spectrum shown in Fig.\ \ref{fig1}(d). The width of the resonance peak in Fig.~\ref{fig1}(d) is much larger than that of the peaks of Fig.~\ref{fig1}(c) even across a frequency domain more than five times as wide, and the spectrum of Fig.\ \ref{fig1}(d) lacks the cross peaks of Fig.~\ref{fig1}(c), indicating an inhomogeneous fine-structure splitting in addition to the inhomogeneity in overall transition frequency. Figure \ref{fig2} shows projections onto the $\nu_t$ axis of rephasing spectrum amplitudes, allowing a more direct comparison of linewidths. As shown by the red trace in Fig.\ \ref{fig2}(a), the PL-detected MDCS measurement exhibits peaks with an inhomogeneous full-width at half-maximum (FWHM) linewidth of $28 \pm 2$ GHz. This is within a factor of two of linewidths obtained through conventional PL [Fig.~\ref{fig2}(a), blue trace] \cite{sm}. Figure \ref{fig2}(b) shows PL-detected MDCS and heterodyne-detected MDCS on the same plot. The heterodyne spectrum's linewidth can be extracted by dividing out the squared laser spectrum and measuring the result to produce an estimate of $1.8 \pm 0.1$ THz, or by fitting to a finite-bandwidth MDCS lineshape model \cite{Smallwood2017} to produce an estimate of $1.84 \pm 0.02$ THz.

As mentioned above, the PL-detected spectrum of Fig.~\ref{fig1}(c) is sensitive only to bright color centers while the heterodyne-detected spectrum of Fig.~\ref{fig1}(d) is sensitive to both bright and hidden centers.The wider distribution of states must be dominated by hidden centers because is not visible in Fig.~\ref{fig1}(c). A comparison between the heterodyne-detected inhomogeneous linewidth and the linewidths from partially annealed and highly-strained \ce{SiV^-} samples reported in the literature \cite{Evans2016,Lindner2018} indicates that the likely origin of the hidden ensemble is strain. Indeed, strain is expected to be nonuniform and potentially very large at selected locations in this densely implanted sample \cite{sm}. The results from Figs.~\ref{fig1} and \ref{fig2} are also similar to a recently reported population of SiV-related optical transitions in nanodiamonds \cite{Lindner2018}, where strain effects are at among their most acute. The discovery here of a significant inhomogeneous population of color centers in single-crystal bulk extends both the phenomenon's relevance and its regime of applicability.

Our assignment of the hidden center resonances to strain-impacted \ce{SiV^-} centers as opposed to some other kind of silicon-related complex is based on the fact that the hidden center resonance is peaked at the \ce{SiV^-} zero-phonon line and---while broader than the bright center resonance---remains too narrow in this sample to originate from any other source. The most plausible alternate contender is the neutral carbon vacancy (GR1) zero-phonon line at 404.5 THz (740.9 nm). However, this resonance lies decidedly away from the hidden-center spectral peak as demonstrated in both Fig.\ \ref{fig2}(b) and in the results of auxiliary MDCS and linear absorption measurements \cite{sm}. The nanodiamond study of Ref.~\cite{Lindner2018} posits the influence of SiV:H complexes. However, SiV:H complexes are expected to absorb in the near-infrared \cite{Thiering2015}, even farther from the hidden-center peak than the GR1 line.

In the absence of other complexes, the clear distinction between the heterodyne-detected and PL-detected \ce{SiV^-} center ensembles begs a microscopic explanation. The simplest possibility is that the presence of strain in the hidden centers reduces their radiative dipole moments relative to unstrained centers, thereby both reducing radiative decay from the excited state and extending its coherence time through the same underlying mechanism. Such an explanation is inconsistent with the nonlinear character of MDCS peak strengths, however, which depend on the dipole moment to the 4th power in the $\chi^{(3)}$ limit. It would be surprising, for example, that the unstrained centers are not observed in Fig.\ \ref{fig1}(d) as localized peaks in the spectrum, since their dipole moments in this picture are supposedly much bigger than the dipole moments of the strained centers. Furthermore, a change in the dipole moment only affects the coherence time in the coherent limit where $T_2 = 2T_1$ \cite{BoydT1T2}. Again, this is inconsistent with our data, as discussed in more detail below.

\begin{figure}[tb]\centering\includegraphics[width=3.4in]{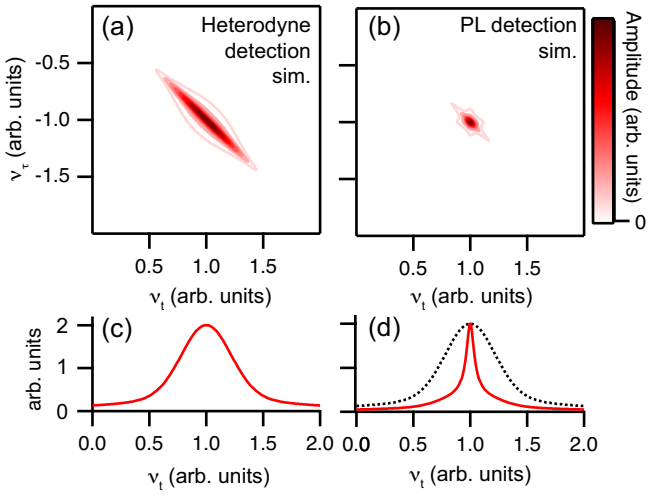}
\caption{\label{fig3}Simulated effect of PL-detection filtering in MDCS \cite{sm}.
{\bf(a)} Heterodyne-detected rephasing plot resulting from an inhomogeneously broadened ensemble of two-level atoms connected to a nonradiative dark state with strain-dependent coupling.
{\bf(b)} PL-detected rephasing plot from the same system. 
{\bf(c)--(d)} Amplitude projections of (a) and (b) onto the $\nu_t$ axis. For comparison, the dashed line in (d) is a copy of the solid red line in (c).
}
\end{figure} 

An explanation of the phenomenology that is more consistent with our results is that strain enhances the coupling between the $^2E_u$ excited states and an electronic dark state in the \ce{SiV^-} center system. For example, the dark state may be shifted by strain from an energy higher than the $^2E_u$ excited states to an energy nearly commensurate with them, or selection rules and phonon-coupling may be modified to make the transition from the $^2E_u$ states into a dark state at lower energy dark state more likely. In either of these situations, the luminescence would be suppressed by the system relaxing into the dark state but the nonlinear optical response would be unaffected, as confirmed by simulations we have performed to illustrate the effect \cite{sm}. Figure \ref{fig3} summarizes one such scenario, showing the effect of a strain-dependent selection rule on an inhomogeneous ensemble of two-level atoms. Although direct experimental confirmation of such a mechanism remains lacking, the relatively low quantum yield of \ce{SiV^-} centers \cite{Bradac2019} supports its existence, and \ce{SiV^-} dark states have been both theoretically predicted \cite{Gali2013,Jahnke2015,Thiering2018} and experimentally reported \cite{Neu2012,Neu2012a}. 

\begin{figure}[tb]\centering\includegraphics[width=3.4in]{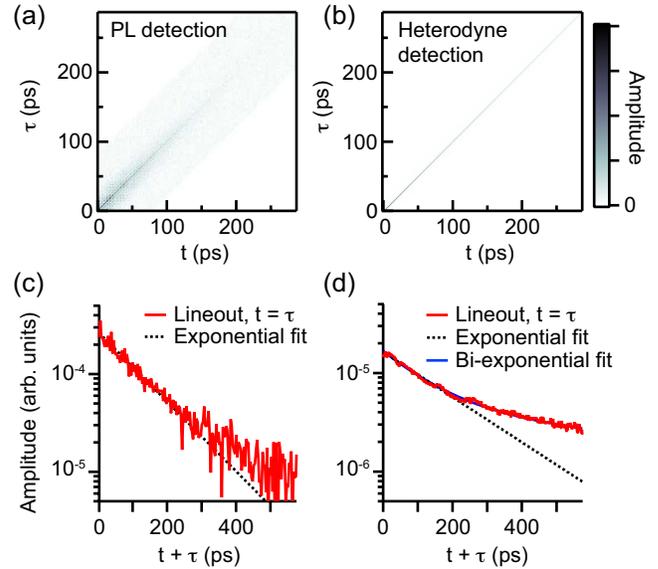}
\caption{\label{fig4} 
Homogeneous dephasing characteristics of the \ce{SiV^-} center ensemble.
{\bf(a)--(b)} Time-domain plots of the data displayed in Figs.\ \ref{fig1}(c) and \ref{fig1}(d).
{\bf(c)--(d)} Diagonal line-outs of the data in (a) and (b) along $\tau = t$, with corresponding exponential fits.
}
\end{figure}

Perhaps the most powerful aspect of MDCS in this study is its ability to extract homogeneous decay rates even in the midst of an inhomogeneous ensemble. Figure \ref{fig4} shows an analysis of these processes, which have been characterized in the time domain to avoid windowing artifacts. Figure \ref{fig4}(a) shows the PL-detected time-domain spectrum used to generate Fig.\ \ref{fig1}(c). Figure \ref{fig4}(b) shows the heterodyne-detected spectrum used to generate Fig.\ \ref{fig1}(d). Figures \ref{fig4}(c) and \ref{fig4}(d) correspond to the amplitudes of the plots in Figs.\ \ref{fig4}(a) and \ref{fig4}(b) extracted along a diagonal lineout at $\tau = t$. In order to provide a numerical estimate of dephasing rates, we fit the data to decaying exponential lineshapes of the form
\begin{equation}
f(t+\tau) = A e^{-(t+\tau)/T_{2a}} + B e^{-(t+\tau)/T_{2b}}, \label{biexp}
\end{equation}
with the second term being omitted for the PL-detected measurement.

As shown by the data and fit in Fig.~\ref{fig4}(c), the PL-detected signature of bright \ce{SiV^-} centers decays with essentially mono-exponential dynamics and exhibits a characteristic homogeneous relaxation time of $T_{2a} = 122 \pm 7$ ps. (The slight deviation from mono-exponential dynamics at times $t+\tau>300$ ps is a manifestation of the experimental noise floor \cite{sm}.) Although shorter than typically reported optical $T_2$ times for isolated \ce{SiV^-} centers \cite{Arend2016,Becker2016,Zhang2017}, this is in line with a recent photon-echo measurement of an \ce{SiV^-} center ensemble \cite{Weinzetl2019}. 

As shown in Fig.~\ref{fig4}(d), the dynamics of the heterodyne-detected \ce{SiV^-} ensemble exhibit a great deal more structure. Although a short-time component similar in scale to the dephasing time of the PL-detected ensemble still exists, the decay curve also exhibits a prominent extended tail that clearly deviates from mono-exponential dynamics. Because the decoherence interactions probed through photon-echo measurements like this are irreversible, the presence of this extended tail is a clear indication that different emitters within the hidden \ce{SiV^-} ensemble exhibit different dephasing rates, with some of these emitters exhibiting $T_2$ times that exceed the $T_2$ times of the PL-detected bright ensemble. This would not be true of measurements like time-resolved PL or transient absorption, where multi-exponential relaxation dynamics have a much less definitive physical interpretation. The application of Eq.~\refeq{biexp} to the data in Fig.~\ref{fig4} leads to extracted coherence times of $T_{2a} = 120 \pm 5$ ps and $T_{2b} = 990 \pm 180$ ps. In the frequency domain, the relaxation dynamics correspond to a bi-Lorentzian lineshape with characteristic linewidths $1/(2\pi T_{2a}) \approx 1.33 \pm 0.06$ GHz and $1/(2\pi T_{2b}) = 160 \pm 30$ MHz. We note, however, that the hidden \ce{SiV^-} center ensemble may include more than just two classes of distinct centers and/or may exhibit a continuous distribution of dephasing times, and so these extracted fit parameters should be interpreted phenomenologically.

Finally, the extracted $T_2$ times can be compared to the sample's electronic population decay times $T_1$, allowing us to verify that the \ce{SiV^-} resonances are in the incoherent limit. We have performed measurements of $T_1$ by measuring the response of our MDCS signal to variations in the waiting time $T$ between the second and third excitation pulses \cite{sm}. The results for both PL detection and heterodyne detection are consistent with time-resolved PL measurements in the literature reporting a lifetime of 1--2 ns \cite{Turukhin1996,Rogers2014b}, under all circumstances much longer than both $T_{2a}/2$ and $T_{2b}/2$.

In conclusion, we have used collinear optical MDCS to measure the coherent properties of an ensemble of \ce{SiV^-} centers in diamond, observing a large and inhomogeneously broadened population of nonradiative electronic states that have remarkably long optical decoherence times in comparison to their radiatively coupled counterparts. The effect can be understood as a likely consequence of strain. Beyond their fundamental relevance to the physics of diamond color centers, the results open opportunities for the controlled use of strain in practical \ce{SiV^-} center applications. One possible use of the effect is in a device in which strain is intentionally applied or modulated in order to controllably tune the amount of radiative emission and electronic coherence in a single-photon emitters or in an emitter ensemble.

\begin{acknowledgments}
We thank G.\ Thiering and E.\ W.\ Martin for useful discussions, and we thank D.\ B.\ Almeida and R.\ C.\ Owen for help with sample preparation. 
R.\ U.\ acknowledges support from a DFG fellowship (UL 474/1-1).
T.\ S.\ acknowledges support from the  Federal Ministry of Education and Research of Germany (BMBF, project DiNOQuant, 13N14921).
Ion implantation work to generate the \ce{SiV^-} centers was performed, in part, at the Center for Integrated Nanotechnologies, an Office of Science User Facility operated for the U.S.\ Department of Energy (DOE) Office of Science. Sandia National Laboratories is a multi-mission laboratory managed and operated by National Technology and Engineering Solutions of Sandia, LLC., a wholly owned subsidiary of Honeywell International, Inc., for the U.S. Department of Energy's National Nuclear Security Administration under contract DE-NA-0003525. This paper describes objective technical results and analysis. Any subjective views or opinions that might be expressed in the paper do not necessarily represent the views of the U.S.\ Department of Energy or the United States Government.
\end{acknowledgments}

\bibliography{References}

\end{document}